\documentclass{IEEEtran4PSCC}

\usepackage{cite}
\usepackage{graphicx}
\usepackage{picture}
\usepackage{multirow}
\usepackage{amsmath}
\usepackage{amsfonts}
\usepackage{bbm}

\usepackage{color}
\usepackage{array}
\usepackage{siunitx}
\usepackage{url}

\usepackage{algorithmic}
\usepackage{algorithm}

\usepackage[caption=false, font=normalsize, labelfont=sf, textfont=sf]{subfig}

\makeatletter
\let\old@ps@headings\ps@headings
\let\old@ps@IEEEtitlepagestyle\ps@IEEEtitlepagestyle
\def\psccfooter#1{%
    \def\ps@headings{%
        \old@ps@headings%
        \def\@oddfoot{\strut\hfill#1\hfill\strut}%
        \def\@evenfoot{\strut\hfill#1\hfill\strut}%
    }%
    \def\ps@IEEEtitlepagestyle{%
        \old@ps@IEEEtitlepagestyle%
        \def\@oddfoot{\strut\hfill#1\hfill\strut}%
        \def\@evenfoot{\strut\hfill#1\hfill\strut}%
    }%
    \ps@headings%
}
\makeatother

\begin{document}

\title{Computation of Price Signals for Controlling Flexible Prosumers in Distribution Grids Incorporating Stochasticity and Fairness}

\author{
\IEEEauthorblockN{Plouton Grammatikos\\ Pierre Roduit \\ Fabrizio Sossan}
\IEEEauthorblockA{School of Engineering \\
HES-SO Valais-Wallis\\
Sion, Switzerland\\
\{plouton.grammatikos, pierre.roduit, fabrizio.sossan\}@hevs.ch}
\and
\IEEEauthorblockN{Chloé Voirol}
\IEEEauthorblockA{School of Engineering \\
EPFL\\
Lausanne, Switzerland\\
chloe.voirol@epfl.ch}
}

\maketitle

\begin{abstract}
%Behind-the-meter distributed energy resources (DERs) have transformed consumers into prosumers, capable of both producing and consuming electricity. A convenient paradigm for coordinating prosumer behavior is through dynamic electricity price signals. By leveraging the flexibility of their DERs, prosumers can respond to these signals to optimize consumption and reduce their energy costs. However, synchronized responses to price signals can lead to grid constraint violations, such as local overproduction or overconsumption.

This paper proposes a method for distribution system operators (DSOs) to design spatially fair dynamic price signals for flexible prosumers, accounting for both distribution grid constraints and the stochastic behavior of prosumers. Since locational pricing can introduce price unfairness when grid constraints are active, the proposed approach includes mechanisms to enforce price fairness across the network.
The method involves the DSO solving a scenario-based optimal power flow (OPF) problem to derive, similarly to locational marginal prices in transmission systems, the Lagrange multipliers associated with grid constraints and to compute individualized price signals for each prosumer under various PV generation and demand scenarios. Subsequently, two methods are introduced to aggregate these individualized prices into a single fair price across prosumers. Their performance is evaluated through simulations on a model distribution grid, comparing prosumer revenues, grid constraint violations, and analyzing the impact of fairness on reducing cost variance among prosumers.
%Behind-the-meter distributed energy resources (DERs) has enabled consumers to transition into prosumers, who can actively participate in generation and demand. By leveraging the DER flexibility, prosumers can react to electricity prices and reduce their bills. However, synchronized reactions may cause grid constraint violations through overproduction or overconsumption. This paper proposes a method for the distribution system operator (DSO) to design price signals that account for grid constraints, prosumers’ stochastic behavior, and fairness. The DSO solves a scenario-based optimal power flow to derive Lagrange multipliers of grid constraints and compute individualized price signals for each prosumer under various PV generation and demand scenarios. Two aggregation methods are then introduced to obtain a single representative price per prosumer. Their performance is compared on a model distribution grid in terms of prosumer revenues and grid violations, and the impact of fairness on reducing cost variance among prosumers is analyzed.
\end{abstract}


\thanksto{\noindent This work was supported by HES-SO under the projects RESINET and EXPOSITION.}

\section{Introduction}
Behind-the-meter distributed energy resources (DERs), such as photovoltaic (PV) panels, battery energy storage systems (BESSs), and electric vehicles (EVs), enable passive consumers to become prosumers. By leveraging the flexibility of their DERs, prosumers can optimize their energy usage and supply valuable services to distribution system operators (DSOs), like congestion management and voltage control, provided that appropriate remuneration is available to do so.

% A convenient paradigm for coordinating prosumer behavior is through dynamic electricity price signals. By leveraging the flexibility of their DERs, prosumers can respond to these signals to optimize consumption and reduce their energy costs. However, synchronized responses to price signals can lead to grid constraint violations, such as local overproduction or overconsumption.
%The growing implementation of behind-the-meter distributed energy resources (DERs), such as photovoltaic (PV) panels, battery energy storage systems (BESS), and electric vehicles (EVs), enables passive consumers to turn into prosumers, which can leverage their flexibility to minimize their electricity bills. Additionally, prosumers can provide their flexibility to the distribution system operator (DSO) to support ancillary services, such as voltage control and congestion management.

Flexibility of behind-the-meter DERs can be activated either directly, through explicit requests to prosumers to change their demand, or indirectly, using price signals \cite{6861959, 6604159}. Typical schemes for computing price signals include time-of-use (TOU) tariffs, which aim to shift demand to off-peak periods (e.g., \cite{9960533, 9733283}), and real-time dynamic tariffs that adapt the price during the day according to the supply and demand \cite{nicolson2018consumer}. Compared to explicit requests, price signals can be easily interpreted by the prosumers as they are provided with incentives to activate their flexibility at a specific time.

TOU and real-time tariffs typically do not consider the distribution grid constraints. As such, they can lead to synchronized prosumer behaviour, with many increasing or reducing demand simultaneously in response to price peaks and troughs, which in turn can cause over- or undervoltages or congestion in the grid's lines. One way to prevent this issue is by using dynamic price signals that account not only for market conditions but also for the physical constraints of the distribution grid. This requires estimating potential grid violations, which inherently depend on the stochastic factors, such as electricity load and PV generation.

In this paper, we propose a method for computing price signals for prosumers that incorporates grid constraints in the prices and accounts for uncertainty in load and PV generation. The goal is to advertise a price signal to each prosumer, such that the prosumers' responses to it respect the grid constraints for every possible realization of the system's uncertainties. We also study the issue of fairness to mitigate locational disparities in the advertised price signals. The robustness and economic performance of the proposed pricing strategy are evaluated, providing insight into the costs the DSO incurs to respect the grid's constraints and fairness among the system's participants.

The price of the distribution's grid constraints is typically estimated in the literature using the Distribution Locational Marginal Prices (DLMPs) method, which utilizes the Lagrange Multipliers of an Optimal Power Flow (OPF) to shift the original electricity price. Linear price signals, as those used in \cite{7524028}, can have multiple optimal solutions and do not guarantee fairness. The works in \cite{6913579, 9061809} extend the notion of DLMPs to define quadratic price signals that are more stable than linear ones, whereas \cite{10418552} addresses the problem by incorporating conic constraints in the Lagrangian. 
Uncertainty can also be incorporated into DLMP computation by solving robust OPFs, as in \cite{8994378, 7835108}, that consider the uncertainty sets of the system's disturbances. However, uncertainty sets fail to capture the time dependence in time series. This can lead to higher costs than necessary, especially when time-dependent constraints are present in the system, as is the case with BESS. In this case, a scenario-based approach is more appropriate, as in \cite{8673657, WANG2025101648}.

Our method handles uncertainty in the load profiles and PV generation by solving a scenario-based quadratic OPF. The DSO computes a price signal for each prosumer and for each scenario by extracting the Lagrange multipliers of the grid constraints for the respective scenario. To avoid multiplicity of solutions, the price signals are computed using the augmented Lagrangian, employed by the Alternating Direction Method of Multipliers (ADMM) \cite{8186925}, which includes a non-linear penalty on the grid's constraints, thereby enhancing the stability of the solution. The works in \cite{6748974, admm-dispatch} use ADMM to solve OPFs in a distributed manner, but do not consider the interpretation of the Lagrangian as a price signal. On the other hand, \cite{8673657, 10402924} utilize the interpretation of the Lagrange multipliers as marginal prices that incentivize the prosumers to change their demand. Our work expands on this interpretation by computing an explicit price signal for each prosumer, allowing them to interpret it outside the context of ADMM.

This paper builds upon our previous work in \cite{Grammatikos_2025}, which interprets the objective of ADMM as a price signal, but does not consider uncertainty.
To account for the worst-case scenario of load and PV generation, the DSO aggregates the price signals computed for each scenario into a robust one by selecting the scenario with the price that differs the most from the original electricity price. The performance of the computed robust price signal is compared to a baseline price signal, which is computed by only considering the expected value of the load trajectories and PV generation.

Fairness is also an issue that arises from the different locations of prosumers in the grid, which is referred to as spatial fairness. Fairness in distribution grids has been studied in the context of PV curtailment (e.g., \cite{GEBBRAN2021116546, 10863114, GUPTA2025126543}). Prosumers located further from the feeder typically experience higher curtailment, as they have a greater impact on nodal voltages. In another context, \cite{8626512} utilizes Jain's fairness index to minimize price disparities among prosumers resulting from DLMPs, while \cite{9565366} balances the costs of prosumers and grid constraints. \cite{WANG2025101648} solves a scenario-based OPF that optimally allocates energy storage among the nodes, considering fairness in the prices experienced by the prosumers. In our work, we consider a fairness objective in the DSO's optimization problem that minimizes the variance of the costs incurred by the prosumers for grid connection. The fairness is then transferred to the computed price signals. We finally perform an analysis of the effect of the degree of fairness on the total costs of the system and the grid constraints violations.

The rest of the paper is organized as follows. Section~\ref{sec:opf_formulation} formulates the centralized scenario-based optimization problem solved by the DSO. Section~\ref{sec:price_signals} described the method for computing the price signals for the prosumers. Section~\ref{sec:validation} evaluates the performance of the computed price signals. Finally, Section~\ref{sec:conclusion} concludes the paper.

\section{Optimization Problem Formulation}
\label{sec:opf_formulation}

\subsection{Use case and notation}
We consider a low-voltage distribution grid interfacing 
$N$ prosumers. Each prosumer $n \in \{1,\dots,N\}$ is equipped with loads, PV panels, and a local BESS. The loads are assumed nonflexible, the PV generation is curtailable (if necessary), and the BESS operates in all four quadrants, providing both active and reactive power within the limits imposed by the power converter and the battery’s characteristics. 

To optimize the use of the BESS energy capacity and maximize the available capacity for local and grid services, each prosumer schedules its operation day-ahead. To this end, we consider $K$ discrete steps $t_k \in \{t_1,\dots, t_K\}$ of the next-day optimization horizon with resolution $\Delta T$, and $S$ day-ahead scenarios $s \in \{1,\dots,S\}$ of different combinations of prosumers' demands and PV generation forecasts.
We denote by $\{p^s_{l,n}(t), q^s_{l,n}(t)\}$, $\{p^s_{pv,n}(t), q^s_{pv,n}(t)\}$, and $\{p^s_{b,n}(t), q^s_{b,n}(t)\}$ the active and reactive power of the loads, the PV generation, and the BESS discharging power (negative if charging) for time step $t$, prosumer $n$, and scenario $s$, respectively.
We define the aggregated nodal demand of the prosumer as the tuple
\begin{align}
\begin{aligned}
x^s_n(t) = [&p^s_{l,n}(t)-p^s_{pv,n}(t)-p^s_{b,n}(t), \\
&q^s_{l,n}(t)-q^s_{pv,n}(t)-q^s_{b,n}(t)].
\end{aligned}
\end{align}
Although the proposed use case assumes the presence of the aforementioned flexible resources, the formulation is general and can accommodate additional resources.

\subsection{Overview of the proposed control-by-price paradigm}
It is assumed that the prosumers share with the DSO the specifications of their assets connected to the grid (as mandated by distribution grid codes in several countries) as well as real-time measurements. However, the DSO cannot directly control the behind-the-meter flexibility, which remains fully under the prosumers’ control. Instead, the DSO can influence prosumers’ consumption patterns through a dynamic price signal.

In particular, the DSO, capitalizing on its full observability, solves a scenario-based centralized OPF and extracts the Lagrange multipliers associated with the active grid constraints of each scenario. These multipliers serve as locational price signals, indicating the marginal cost of enforcing network constraints. To ensure usability, each prosumer receives a single, time-varying price signal. This is derived by aggregating the scenario-specific multipliers into one representative signal per prosumer. Finally, the prosumer optimizes its own energy consumption, driven by the new price signal and unaware of any grid constraints.

The rest of this section presents the centralized optimization problem that the DSO must solve. The method for computing the price signals is presented in Section~\ref{sec:price_signals}.

\subsection{Prosumer Objective}
The goal of the prosumers is to minimize their electricity bill. Given a forecast of the retail electricity price $\mathbf{c}=[c_{t_1}, 0, \dots c_{t_K}, 0]^\top$, where the zeros correspond to the price of reactive power, the forecast electricity cost paid by prosumer $n$ for scenario $s$ is the following
\begin{equation}
    \label{eq:prosumer_cost}
    C_{n}^s = \mathbf{c}^\top \mathbf{x}_n^s
\end{equation}
where $\mathbf{x}_n^s=[x_n^s(t_1), \dots, x_n^s(t_K)]^\top$ is the collection of the aggregated demand for all time instances.

\subsection{PV Constraints}
\begin{subequations} \label{eq:pv_constraints}
We denote by $G^s_n(t) \in [0,1]$ the per-unit forecasts of PV generation at the location of prosumer $n$ for time $t$ and scenario $s$.
The active powers of the PV panels are bounded by the following constraints  
\begin{equation}
    0 \leq p^s_{pv,n}(t) \leq p_{pv,n}^{max} G_n^s(t)
\end{equation}
where $p_{pv,n}^{max}$ is the maximum power of the PV of prosumer $n$. For simplicity, we assume that the PV panels only offer active-power flexibility, so we consider
\begin{equation}
    q^{s}_{pv,n}(t) = 0, \forall n,t,s
\end{equation}
\end{subequations}

\subsection{BESS Constraints}

We use a box approximation of the PQ capability curve of the BESS's converter in the four quadrants:
\begin{subequations} \label{eq:bess_constraints}
\label{eqs:battery}
    \begin{align}
    -s_{b,n}^{max} &\leq \sqrt{2} \cdot  p^s_{b,n}(t) \leq s_{b,n}^{max} \\
    -s_{b,n}^{max} &\leq \sqrt{2} \cdot q^s_{b,n}(t) \leq s_{b,n}^{max}
\end{align}
where $s_{b,n}^{max}$ is the maximum apparent power of the converter.

The state of charge (SOC) of the BESS is defined as a time coupling constraint, which connects the current BESS capacity to its power history. The SOC is also limited to ensure safe operations and to enhance battery life, within a range $[SoC^{min}_n, SoC^{max}_n] = [10\%, 90\%]$. The SOC with unitary efficiency model reads as:
\label{eq:bess}
\begin{align}
    \label{eq:bess:soc}
    & SoC^s_n(t) = SoC_{n}(t_1) - \frac{\Delta T}{E_{b,n}} \sum_{\tau=t_1}^{t-1} p^s_{b,n}(\tau) \\
    & SoC^{min}_n \leq SoC^s_{n}(t) \leq SoC^{max}_n
\end{align}
where $SoC_{n}(t_1)$ is the initial SoC for posumer $n$ (same for all scenarios) and $E_{b,n}$ is the energy capacity of the BESS. Non-unity efficiencies can be modeled using relaxations or mixed-integer formulations with binary variables, as is well established in the literature. Since their inclusion is straightforward and does not affect our main contributions, we omit them here for clarity of exposition. 
\end{subequations}

\subsection{Grid Constraints}

\begin{subequations}
The goal of the DSO is to maintain all the nodal voltages within defined limits $[v^{min}, v^{max}]$ and to respect the current limits at all branches and power constraints at the grid connection point (GCP). We denote by $v^s_j(t), j \in \{1,\dots,M\}$ the voltage at node $j$, at time $t$ for scenario $s$, where $M$ is the number of nodes in the grid.
The relation between nodal power injections and voltages is expressed using the following linearized approximation
\begin{equation}
\begin{aligned}
    \label{eq:lin_grid}
    v^s_j(t)=v^{s*}_j&+K^s_{v,p}(t)(\mathbf{p}^s(t)-\mathbf{p}^{s*}(t))+\\
    &+K^s_{v,q}(t)(\mathbf{q}^s(t)-\mathbf{q}^{s*}(t)),
\end{aligned}
\end{equation}
where $\mathbf{p}^s(t)$ and $\mathbf{q}^s(t)$ are the vectors of all the nodal active and reactive power injections, and $\mathbf{p}^{s*}(t), \mathbf{q}^{s*}(t)$ and $v^{s*}_l$ are the injections and voltages at the grid linearization point, respectively. The latter is computed with a load flow at the reference power injections; $K^s_{v,p}(t), K^s_{v,q}(t)$ are the matrices of the sensitivity coefficients of the voltages with respect to the active and reactive powers, computed around the reference points. The sensitivity coefficients are the derivatives of the nodal voltages with respect to the active and reactive nodal power injections and can be computed analytically \cite{sensitivities}.
The linearized voltage constraints can then be expressed by
\begin{equation}
    v^{min} \leq v_j^s(t) \leq v^{max}.
\end{equation}

Similarly, we define the constraints of the linearized branch current magnitudes $i_l^s(t), l \in \{1,\dots,L\}$, where $L$ is the number of branches in the grid, and the linearized slack's active and reactive power $p_0^s(t), q_0^s(t)$. The linear approximations for the branch currents and the slack's power are given by equations similar to \eqref{eq:lin_grid}. The current magnitude constraints can be written as
\begin{equation}
    0 \leq i_l^s(t) \leq i_l^{max},
\end{equation}
where $i_l^{max}$ is the ampacity rating of branch $l$.
% Similarly, we can compute a linear approximation of the active and reactive power at the slack bus $p^s_0(t), q^s_0(t)$ as follows
% \begin{equation}
% \begin{aligned}
%     p^s_0(t)=p^{s*}_0&+K^s_{p_0,p}(t)(\mathbf{p}^s(t)-\mathbf{p}^{s*}(t))+\\
%     &+K^s_{p_0,q}(t)(\mathbf{q}^s(t)-\mathbf{q}^{s*}(t))
% \end{aligned}
% \end{equation}
% \begin{equation}
% \begin{aligned}
%     q^s_0(t)=q^{s*}_0&+K^s_{q_0,p}(t)(\mathbf{p}^s(t)-\mathbf{p}^{s*}(t))+\\
%     &+K^s_{q_0,q}(t)(\mathbf{q}^s(t)-\mathbf{q}^{s*}(t))
% \end{aligned}
% \end{equation}
% where $K^s_{p_0,p}(t), K^s_{p_0,q}(t)$ are the matrices of the sensitivity coefficients of the slack's active power with respect to the nodal active and reactive power injections, and $K^s_{q_0,p}(t), K^s_{q_0,q}(t)$ are the sensitivity coefficients of the slack's reactive power.
The power constraints at the slack bus are typically expressed as a power factor constraint. However, the power factor definition is non-convex, even after linearizing the power flow equations. To avoid non-convexity, we use a tight upper bound $q_0^{max}$ for the slack's reactive power
\begin{equation}
    q^s_0(t) \leq q_0^{max},
\end{equation}
where $q_0^{max} \ll s_0^{max}$ and $s_0^{max}$ is the transformer's maximum apparent power at the GCP. Then, the active power limit is defined as
\begin{equation}
    p^s_0(t) \leq \sqrt{(s_0^{max})^2-(q_0^{max})^2}
\end{equation}
\end{subequations}

In the following, we group all the DSO's constraints into the following equations
\begin{subequations}\label{eq:linear_grid_constraint}
    \begin{align}
        A^{eq,s} \mathbf{X}^s &= \mathbf{b}^{eq,s} \\
        A^{ineq,s} \mathbf{X}^s &\leq \mathbf{b}^{ineq,s},
    \end{align}
\end{subequations}
where $\mathbf{X}^s$ is the matrix of all the prosumers' aggregated power injections $\mathbf{x}^s_n$ for scenario $s$; $A^{eq,s}, A^{ineq,s}$ are the equality and inequality linearization matrices, and $\mathbf{b}^{eq,s}, \mathbf{b}^{ineq,s}$ are the linearization vectors of the grid constraints for scenario $s$. This formulation will be useful in computing the price signals in Section~\ref{sec:price_signals}.

\subsection{Centralized Problem's Objective}
The problem objective is to minimize the prosumers' total energy bills while respecting the distribution grid's constraints. It is assumed that the DSO has no monetary objective of its own, but may consider fairness, as developed later.

The expected monetary cost of the system is the total cost of the prosumers, averaged over all scenarios:
\begin{equation}
    \mathcal{J}^c = \frac{1}{S} \sum_{s=1}^S \sum_{n=1}^N \mathbf{c}^T \mathbf{x}_n^s.
\end{equation}

Due to the grid constraints, the costs incurred by prosumers will be higher compared to the unconstrained case. We measure fairness by the variance of the cost increases among prosumers compared to the unconstrained case. We denote by $\hat{\mathbf{p}}_n^s$ the vector of the active powers $\hat{p}_n^s(t)$ that minimize the objective of prosumer $n$ \eqref{eq:prosumer_cost} subject to the prosumer's constraints \eqref{eq:pv_constraints} and \eqref{eq:bess_constraints} for scenario $s$. Then, the cost increase of the prosumer at time $t$ is defined as
\begin{equation}
    \Delta C_n^s (t) = c_{t} (p_n^s(t) - \hat{p}_n^s(t)).
\end{equation}
Then, the fairness objective of the DSO is defined as
\begin{equation}
    \mathcal{J}^f = \frac{1}{S}\sum_{s=1, t=t_1}^{S, t_K} \text{Var} \left( \Delta C_n^s(t) \right),
\end{equation}
where $\text{Var}(\cdot)$ is the sample variance among the prosumers.

\subsection{Centralized Problem Formulation}

% Given $\tilde{\mathbf{x}}^s_n = [\mathbf{p}^s_{b,n}, \mathbf{q}^s_{b,n}, \mathbf{p}^s_{PV,n}, \mathbf{q}^s_{PV,n}, \mathbf{soc}^s_n]$, $\mathbf{x}^s_n$ the vector of the aggregated demands $x^s_n(t)$ and.

The vector of the optimization variables for all timesteps of prosumer $n$ and scenario $s$ is
\begin{equation}
    \tilde{\mathbf{x}}^s_n = [\mathbf{p}^s_{b,n}; \mathbf{q}^s_{b,n}; \mathbf{p}^s_{pv,n}; \mathbf{q}^s_{pv,n}; \mathbf{soc}^s_n]
    \end{equation}
where $[\cdot;\cdot]$ denotes vertical concatenation of vectors, or matrices. The centralized OPF is formulated as follows
\begin{subequations}
    \label{eq:centralized_problem}
    \begin{align}
    \label{eq:objective}
    & \min_{\mathbf{x}^s_n, \tilde{\mathbf{x}}^s_n} && \mathcal{J}^c + w^f \mathcal{J}^f + \alpha \frac{1}{S} \sum_{n=1, s=1}^{N,S} \|\mathbf{p}_{b,n}^s\|_2^2 \\
    &\text{subject to: } && 
    \text{PV constraints } \eqref{eq:pv_constraints} \\
    & &&\text{BESS constraints } \eqref{eq:bess_constraints} \\
    & &&\text{Linear grid constraints } \eqref{eq:linear_grid_constraint} \\
    &&&\forall n \in \{1,\dots,N\}, s \in \{1,\dots,S\},
    \end{align}
\end{subequations}
% where $\mathbf{c}$ is the day-ahead spot-market electricity price (typically 0 for the reactive power), $\mathbf{f}_n^{eq}, \mathbf{f}_n^{ineq}$ represent the equality and inequality constraints of load, BESS, and generating units of prosumer $n$, and $\mathbf{g}^{eq}, \mathbf{g}^{ineq}$ are the grid equality and inequality constraints. 
where $w^f$ is the weight of the fairness objective, and the terms $\|\mathbf{p}_{b,n}^s\|_2^2$ with weight $\alpha$ are small penalties on the BESS power that prevent the multiplicity of solutions inherent to linear problems. Problem~\eqref{eq:centralized_problem} is convex, because the variance and norm are convex and all the constraints are linear.

% To have a convex problem, the functions $\mathbf{f}_n^{eq}$ must be linear and the functions $\mathbf{f}_n^{ineq}$ must be convex. This can be enforced by design. However, $\mathbf{g}^{eq}, \mathbf{g}^{ineq}$ are inherently non-convex, because of the non-convex characteristics of the power flow equation. Therefore, we use a linearized approximated optimal power flow. The grid constraints are then expressed as follows:

To minimize the error of the linearized model, we recompute the linearization matrices until the linearization error of the grid constraints is below a given tolerance $\epsilon$. We aggregate the matrices $A^{eq,s}, A^{ineq,s}, \mathbf{b}^{eq,s}, \mathbf{b}^{ineq,s}, \mathbf{X}^s$, into matrices $A^{eq}, A^{ineq}, \mathbf{b}^{eq}, \mathbf{b}^{ineq}, \mathbf{X}$ containing the respective matrices for all scenarios. We also define
\begin{subequations}
    \begin{align}
        % A^{eq} = [A^{eq,1};\dots; A^{eq,S}]\\
        % A^{ineq} = [A^{ineq,1};\dots; A^{ineq,S}]\\
        A = [A^{ineq};A^{eq}] \\
        % \mathbf{b}^{eq} = [\mathbf{b}^{eq,1};\dots; \mathbf{b}^{eq,S}] \\
        % \mathbf{b}^{ineq } = [\mathbf{b}^{ineq,1}; \dots; \mathbf{b}^{ineq,S}] \\
        \mathbf{b} = [\mathbf{b}^{ineq};\mathbf{b}^{eq}].
        % \\
        % \mathbf{X} = [\mathbf{X}^1;\dots;\mathbf{X}^S],
    \end{align}
\end{subequations}
We denote by $\mathbf{g}^{eq}(\mathbf{X})$, $\mathbf{g}^{ineq}(\mathbf{X})$ the vectors of the non-convex inequality and equality grid constraints evaluated at the set of setpoints $\mathbf{X}$, such that $\mathbf{X}$ are feasible if $\mathbf{g}^{eq}(\mathbf{X}) = 0$ and $\mathbf{g}^{eq}(\mathbf{X}) \leq 0$. Then, the linearization error of the grid constraints is defined by the norm
\begin{equation}
    \|A\mathbf{X}-\mathbf{b}-\mathbf{g}(\mathbf{X})\|_2,
\end{equation}
where $\mathbf{g}(\mathbf{X}) = [\mathbf{g}^{eq}(\mathbf{X}); \mathbf{g}^{ineq}(\mathbf{X})]$.

The algorithm for approximating the solution of the non-convex OPF by sequentially solving a series of linearised OPFs is summarized in Algorithm~\ref{alg:sequential_linearization}, where the superscript $^k$ is used to indicate the iteration number.
It should be noted that \eqref{eq:centralized_problem} is separable with respect to the scenarios so that we can solve $S$ simpler problems, one for each scenario.

\begin{algorithm}
\caption{Sequential Linearization of OPF}
\label{alg:sequential_linearization}
\begin{algorithmic}[1]   % [1] makes the lines numbered
\STATE \textbf{Input:} load and PV generation scenarios $p_{l,n}^s(t)$, $q_{l,n}^s(t)$, $G_n^s(t)$. Grid topology. Convergence tolerance $\epsilon$.
\STATE \textbf{Output:} aggregated demands $\mathbf{x}^s_n$, optimal resources setpoints $\tilde{\mathbf{x}}^s_n$, constraint matrices $A$ and $\mathbf{b}$, Lagrange multipliers $\lambda$
\STATE $k \gets 1$
\STATE initialize $\mathbf{X}^k, A^k, \mathbf{b}^k$
\WHILE{$\|A^k\mathbf{X}^k-\mathbf{b}^k-\mathbf{g}(\mathbf{X}^k)\|_2 > \epsilon$}
    \STATE Solve \eqref{eq:centralized_problem} with\\ $(A^{eq,s})^k(\mathbf{x})=(\mathbf{b}^{eq,s})^k, \forall s$\\
    $(A^{ineq,s})^k(\mathbf{x})=(\mathbf{b}^{ineq,s})^k, \forall s$, \\to get the optimal demands $(\mathbf{x}^s_n)^k$, resources setpoints $(\tilde{\mathbf{x}}^s_n)^k$ and Lagrange multipliers $\lambda^k$ and compute the nodal power injections $\mathbf{X}^{k+1}$
    \STATE Recompute $A^k, \mathbf{b}^k$ around $\mathbf{X}^{k+1}$
    \STATE $k \gets k+1$
\ENDWHILE
 \STATE \textbf{return} $(\mathbf{x}^s_n)^k, (\tilde{\mathbf{x}}^s_n)^k, A^k, \mathbf{b}^k, \lambda^k$
\end{algorithmic}
\end{algorithm}

\section{Price Signal Computation}
\label{sec:price_signals}

\subsection{Lagrangian}

Instead of advertising the resources' setpoints $\tilde{\mathbf{x}}^s_n$ directly to the prosumers, the DSO computes a price signal for the aggregated demand $\mathbf{x}_n$. These price signals are computed such that they represent, in addition to the original price, potential grid-constraints violations. Suppose the prosumers optimize their operation given the original price $\mathbf{c}$. In that case, their decisions might violate the grid's constraints, because the prosumers' behaviours will be synchronized according to the dynamics of $\mathbf{c}$. Therefore, the goal of the DSO is to shift the optimal operation schedule of each prosumer towards one that allows for safe grid operations by advertising separate price signals to each one.

% In cases where the individual optimal schedules \eqref{eq:indiv_problem} would lead to grid constraints violations, this process aims to shift the optimal operation schedule towards one that allows for safe grid operations. 

We formulate these price signals using the augmented Lagrangian for the grid constraints of the linearized centralized problem \eqref{eq:centralized_problem} defined in \cite{8186925}. The formulation follows a similar methodology as in \cite{Grammatikos_2025}. The Augmented Lagrangian of the problem moves the coupling constraints of \eqref{eq:linear_grid_constraint} to the cost function as follows
\begin{equation} 
    \label{eq:lagrangian_centr}
    \begin{aligned}  
        \mathcal{L}(\mathbf{X}) = &\mathcal{J}^c + w^f \mathcal{J}^f + \mathbf{\lambda}^\top(A\mathbf{X}-\mathbf{b})
        +\frac{\rho}{2}||A^{eq}\mathbf{X}-\mathbf{b}^{eq}||_2^2 +\\&+ \frac{\rho}{2}||\max(A^{ineq}\mathbf{X}-\mathbf{b}^{ineq}, 0)||_2^2.
    \end{aligned}  
\end{equation}
The first two terms represent the original cost of the system. The penalty on the BESSs power is omitted, and instead is incorporated into the prosumers' optimization problems (see Section~\ref{sec:indiv_problem}). The third term attempts to enforce the feasibility of the dual problem. The last two terms are additional quadratic penalties on the constraint satisfaction with weight $\rho$ that help with stability. Minimizing \eqref{eq:lagrangian_centr} subject to the prosumers' constraints \eqref{eq:pv_constraints} and \eqref{eq:bess_constraints} should give the solution to the centralized problem \eqref{eq:centralized_problem}. Therefore, it can be perceived as a price signal that enforces the grid constraints.

However, \eqref{eq:lagrangian_centr} includes some coupling terms, so it cannot be used to extract separate price signals. The cost $\mathcal{J}^c$ is separable among the prosumers and can be written as a sum of the individual monetary costs
\begin{equation}
    \mathcal{J}_n^c = \frac{1}{S} \sum_{s=1}^S \mathbf{c}^\top \mathbf{x}_n^s.
\end{equation}
On the other hand, the fairness penalty $\mathcal{J}^f$ cannot be separated because of coupling terms in the variance. To split the costs, we use an approximation of the variance of the cost increases as follows
\begin{equation}
    \label{eq:var_approx}
    \text{Var}(\Delta C_n^s(t)) \approx \frac{1}{N} \sum_{n=1}^N \left( \Delta C_n^s(t) - \mu^s(t) \right)^2,
\end{equation}
where 
\begin{equation}
    \mu^s(t) = \frac{1}{N} \sum_{n=1}^N \bar{\Delta C_n^s}(t),  
\end{equation}
and $\bar{\Delta C_n^s}(t)$ is the value of $\Delta C_n^s(t)$ computed by solving the centralized problem \eqref{eq:centralized_problem} for scenario $s$. Therefore, we approximate the expected value of the prosumers' costs as the average of the optimal costs over all prosumers. \eqref{eq:var_approx} is now separable among prosumers, because $\mu^s(t)$ is a constant (it does not depend on $\Delta C_n^s(t)$).

To separate the last two terms of \eqref{eq:lagrangian_centr}, we use ideas inspired by the method of dual decomposition \cite{8186925}. First, we partition the matrix $A$ column-wise as follows
\begin{equation}
     A=\begin{bmatrix} A_1 & \dots & A_N \end{bmatrix},
\end{equation}
where $A_n$ contains the columns of A corresponding to prosumer $n$. We also introduce the auxiliary variables: 
\begin{equation}
    \mathbf{z}_n^s = A_n^s \bar{\mathbf{x}}^s_n, 
\end{equation}
with $\bar{\mathbf{x}}^s_n$ being the solution of the centralized problem for prosumer $n$ and scenario $s$ and $A_n^s$ contains the rows of $A_n$ that correspond to the grid constraints tied with scenario $s$.

Then, the augmented Lagrangian of prosumer $n$ and scenario $s$ can be defined as:
\begin{align}
\mathcal{L}_n^s(\mathbf{x}^s_n) = &\mathbf{c}^\top \mathbf{x}^s_n + \frac{w^f}{N} \|\mathbf{c}^\top (\mathbf{x}_n^s-\hat{\mathbf{x}}_n^s)-\mu^s\|_2^2+\\
&+(\mathbf{\lambda}^s_n)^\top(A^s_n\mathbf{x}^s_n-\mathbf{z}^s_n)+\frac{\rho}{2}||A^s_n\mathbf{x}^s_n-\mathbf{z}^s_n||_2^2,
\end{align}
where $\hat{\mathbf{x}}_n^s$ are the solutions of the unconstrained problems and $\mu^s$ is the vector of $\mu^s(t)$. 
% As such, this formulation penalizes deviations from the optimal value of the centralized problem, which is the optimal value within the solution space defined by the grid constraints. 
Reformulation of this augmented Lagrangian highlights the quadratic nature of the cost.
\begin{equation} \label{eq:lagrangian_rewrite}
\begin{aligned}
&\mathcal{L}_n^s(\mathbf{x}_n^s) = \\
&\left[\mathbf{c} \odot \left(\mathbbm{1} - \frac{w^f}{N}\mu^s \right) + (A_n^s)^\top \lambda_n^s - \rho (A_n^s)^\top \mathbf{z}^s_n \right]^\top \mathbf{x}_n^s + \\
& + (\mathbf{x}_n^s)^\top \left[ \frac{\rho}{2}  (A_n^s)^\top A_n^s + \frac{w^f}{N} \text{diag}(\mathbf{c}) \right] \mathbf{x}_n^s + \text{const},
\end{aligned}
\end{equation}
where $\mathbbm{1}$ is a vector with all elements $1$, $\odot$ indicates element-wise multiplication of vectors and $\text{diag}(\mathbf{c})$ is a diagonal matrix with the elements of $\mathbf{c}$ in the diagonal. Expression~\eqref{eq:lagrangian_rewrite} is in the standard form of the cost function of a quadratic program:
\begin{equation}
    \mathcal{L}_n^s(\mathbf{x}_n^s) = (\tilde{\mathbf{c}}_n^s)^\top \mathbf{x}_n^s + \frac{1}{2} (\mathbf{x}_n^s)^\top Q_n^s \mathbf{x}_n^s + \tilde{C}_n^s,
\end{equation}
where $\tilde{\mathbf{c}}_n^s$ is the modified price that considers the grid constraints and fairness, $Q_n^s$ is a price on the square of the demand, which helps with stability and fairness, and $\tilde{C}_n^s$ is a constant cost that can be interpreted as a fee for grid connection.

The stochastic variables introduce uncertainty in the price signal computation, resulting in multiple Lagrangians, one for each scenario. However, the DSO should advertise only one price signal to the prosumers.
% However, the uncertainty should not be propagated to the price signal, because the uncertainty in the demand of one prosumer might affect the price signal of another one (e.g., through the matrices $A_n^s$); therefore, only one price signal should be transmitted to the prosumer. 
We compare two approaches to aggregate the Lagrangians into a single representative price signal. 

\subsubsection{Method 1: expected scenarios}
The augmented Lagrangian is computed using the average load and PV generation among all scenarios. For this method, there is only one scenario in \eqref{eq:centralized_problem}, resulting in a single Lagrangian for each prosumer.

\subsubsection{Method 2: robust price signal}
The expected scenario may result in grid-constraints violations because it does not account for the worst-case scenario. In this method, we aim to compute a worst-case price signal to ensure the grid constraints are respected for every possible scenario of the load and PV generation. The computation relies on the idea that the modified price $\tilde{\mathbf{c}}_n^s$ will differ from the original price $\mathbf{c}$ when there are binding constraints in the system. Therefore, for each timestep, we select the coefficients of the Lagrangian for the scenario whose linear price signal component is most different from the original price. Formally, the robust price signal $t$ is given as follows
\begin{align}
    &\mathcal{L}_n^{robust}(t_{2n-1}) = \{\mathcal{L}_n^s(t_{2n-1}) : \min_s |\tilde{c}_n^s(t_{2n-1}) - c_{t_n}|\} \\
    &\mathcal{L}_n^{robust}(t_{2n}) = \{\mathcal{L}_n^s(t_{2n}) : \min_s |\tilde{c}_n^s(t_{2n})|\},
\end{align}
where the elements of the price signal with odd indices correspond to the active power coefficients and the even ones correspond to the reactive power coefficients (the original price for the reactive power is 0).
% It aims to compensate at each time step for the scenario that represents the most significant risk for grid violations. We identify the scenario with the most considerable violation risk as the scenario whose linear price signal component is most different from the original price (${\Delta}(t) = | \tilde{c}(t) - c(t)|$). This method offers the advantage of being computed offline and with no interaction with the prosumer.
The two methods are compared in terms of achieved costs, constraints violations, and fairness in Section~\ref{sec:validation}.

Finally, we should note that the above formulation of price signals is general and can be applied to additional types of resources and grid constraints beyond those defined in Section~\ref{sec:opf_formulation}. The only limitation is that the grid model must be linear, so that the grid constraints are separable.

\subsection{Individual problem formulation} \label{sec:indiv_problem}
Using one of the two proposed methods, the DSO computes one price signal per prosumer of a quadratic nature.
The linear, quadratic, and constant parts for prosumer $n$ are respectively expressed by $\tilde{\mathbf{c}}_n$, $Q_n$, and $\tilde{C}_n$. Then, the prosumers schedule their demand by minimizing this cost independently and unaware of grid constraints. The optimization problem to be solved by each prosumer is the following
% \todo{Is this problem solved for every scenario? What does the prosumer implement in the end? Shouldn't each prosumer also solve a scenario-based optimization? They would have fewer scenarios and compute a single BESS and PV trajectory.}
\begin{subequations}
    \label{eq:indiv_problem}
    \begin{align}
    \label{eq:objective}
    & \min_{\tilde{\mathbf{x}}_{n}, \mathbf{x}_n}  &&  \tilde{\mathbf{c}_n}^\top \mathbf{x}_n + \frac{1}{2}\mathbf{x}_n^\top Q_n \mathbf{x}_n + \tilde{C}_n + \alpha \|\mathbf{p}_{b,n}\|_2^2\\
    &\text{subject to: } && 
    \text{PV constraints } \eqref{eq:pv_constraints} \\
    & &&\text{BESS constraints } \eqref{eq:bess_constraints},
    \end{align}
\end{subequations}
where the prosumer only considers a single scenario of the load and PV generation. In the following section, we examine the outcome when prosumers solve \eqref{eq:indiv_problem} for various combinations of scenarios. We should note that the prosumers themselves could solve a scenario-based optimization problem to determine their scheduled demand. This could be studied in future research.

\section{Validation} \label{sec:validation}

\subsection{Case Study}

\begin{figure}[t!]
    \centering
    \includegraphics[width=\linewidth]{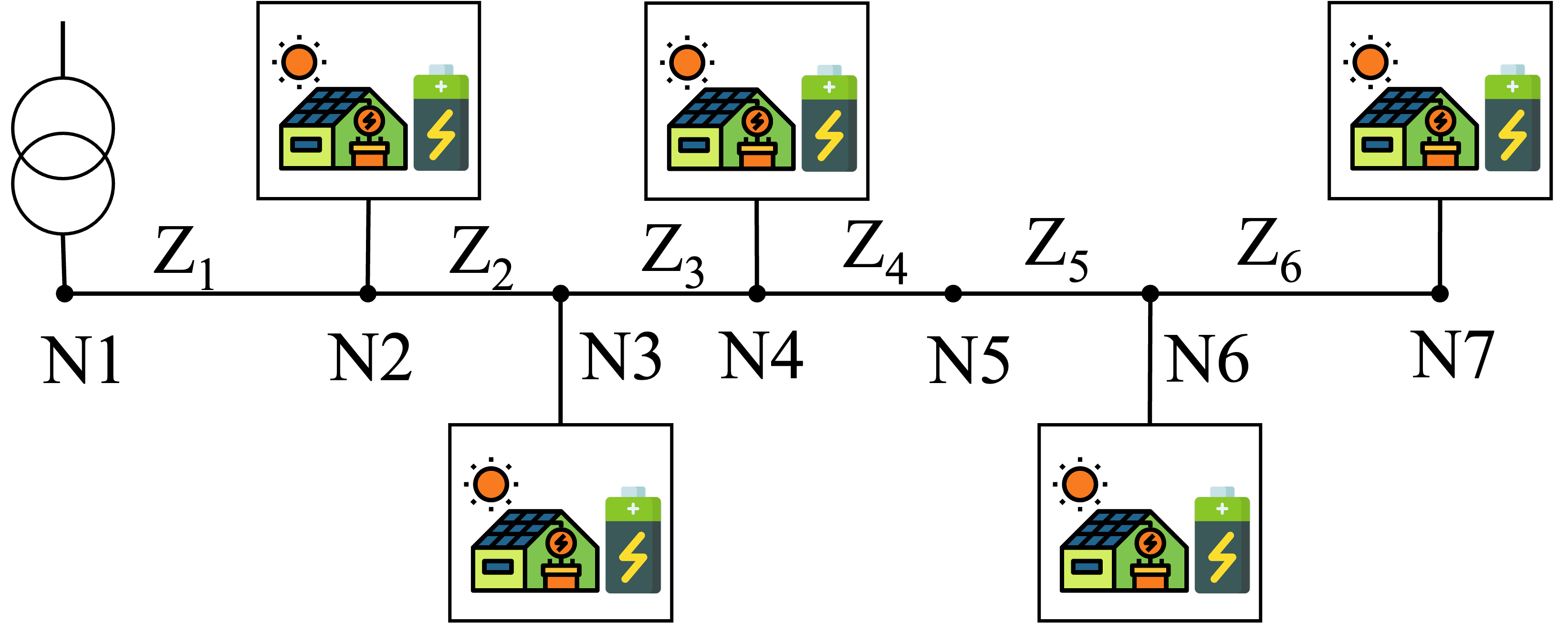}
    \caption{Grid topology for the experimental validation. N1 is the connection point with the upper grid.}
    \label{fig:topology}
\end{figure}

\begin{table}[t!]
    \caption{Power and Energy Ratings of Prosumers}
    \renewcommand{\arraystretch}{1.3}
    \centering
    \begin{tabular}{l|l}
         \bf Parameter & \bf Value \\
         \hline
         Max Load & 2.5 kW \\
         Max PV power & 6 kW \\
         Rated BESS power & 4 kVA \\
         BESS energy capacity & 4 kWh\\
    \end{tabular}
    \label{tab:prosumer_ratings}
\end{table}

\begin{table}[t!]
    \renewcommand{\arraystretch}{1.3}
    \caption{Line parameters}
    \centering
    \begin{tabular}{c|c|c}
         \bf Line & \bf Resistance (m$\Omega$) & \bf Reactance (m$\Omega$)  \\
         \hline
         $Z_1$ & 585.1 & 305.7\\
         $Z_2$ & 194.1 & 161.3\\
         $Z_3$ & 216.1 & 177.8\\
         $Z_4$ & 200 & 0 \\
         $Z_5$ & 597.0 & 296.2\\
         $Z_6$ & 194.4 & 166.4
    \end{tabular}
    \label{tab:line_parameters}
\end{table}

We validate the proposed price signals via simulations on a low-voltage, three-phase, radial distribution grid implemented in the Gridlab laboratory at HES-SO Valais in Sion (Switzerland), as shown in Fig.~\ref{fig:topology}. Node N1 is connected to the external grid and serves as the system's slack bus. Five prosumers are connected to nodes N2, N3, N4, N6, and N7, each one owning a curtailable PV and a BESS. The power and energy rating of the prosumers' assets are the same and are shown in Table~\ref{tab:prosumer_ratings}. The line impedances, shown in Table~\ref{tab:line_parameters}, are estimated from measurements of the experimental setup as described in \cite{11180578}.
%In the experimental setup on which the model is based, impedance $Z_4$ represents a real cable, whereas the other impedances are emulated to model aerial lines of length 500m-1km; the line parameters are estimated using a numerical procedure based on measurements . Since our methodology assumes a single phase, we use a single-phase balanced equivalent of the three-phase network. 

The voltage limits of the system are set to [0.9, 1.05]pu. The transformer's rating at the GCP is $15$~kVA. The maximum reactive power at the slack bus is set to $q_0^{max} = 0.1s_0^{max}$, which gives a power factor of over 0.99 when the exported power to the main grid is maximum. Because we focus the performance evaluation on potential voltage-constraints violations, we set very high limits for the ampacity limits of the grid's lines. The simulations are performed in MATLAB, and the optimization problems are solved with Gurobi. Because this study focuses on price control and fairness, the grid is assumed balanced and is modeled using a single-phase equivalent.

\subsection{Scenarios Definitions}

% We consider 36 weekdays of real measurements of the electrical load (active and reactive power) in office buildings, taken at a 10-minute resolution. 
We consider 10 actual day-ahead forecasts (2 for each prosumer) of the active and reactive electrical load of office buildings on a weekday. These forecasts have a 10-minute resolution and are characterised by real prediction errors.
% We choose two days for each prosumer to use as their day-ahead forecast load. 
We also consider two normalized PV-generation profiles scaled such that their maximum values are 1 and 0.7, respectively. The normalized PV profiles are the same for all prosumers. We then generate $S=64$ scenarios by considering all possible combinations of the scenarios for the five prosumers' loads and the two scenarios for the PV generation. These scenarios, referred to as ``training scenarios", will be used by the DSO to solve the centralized optimization problem and compute the price signals according to the method of Sections~\ref{sec:opf_formulation} and \ref{sec:price_signals}.

We also consider two load scenarios for each prosumer, sampled from the same distribution as the training ones, but not used to compute the price signals. These scenarios, referred to as ``test scenarios", are used to solve the individual problems \eqref{eq:indiv_problem} to assess whether the computed price signals can respect the grid constraints in cases that the optimizer has not seen. The PV scenarios of the test cases are identical to those in the training scenarios.

\subsection{Evaluation Method}
The performance of the computed price signals is evaluated in terms of the costs experienced by the prosumers and the size of the potential voltage-constraints violations. The evaluation procedure is as follows: We first use the 64 training scenarios to solve the prosumers' individual optimization problems without any grid constraints (base case). We then use the same scenarios to solve the DSO's optimization problem, computing the 64 price signals, which are then aggregated into the robust price signal. We also use the average of the scenarios to compute the expected price signal. We solve \eqref{eq:indiv_problem} independently for each prosumer, using both the expected and robust price signals for each of the training and test scenarios. We finally use the computed aggregated demands $\mathbf{x}_n$ to calculate the expected cost incurred by each prosumer for each scenario, as well as the expected voltage-constraint violations resulting from the scheduled demands.

The performance of the expected and robust price signals is evaluated independently on the training and test scenarios using the following metrics:
\begin{enumerate}
    \item Total cost increase in $\$$ of the system compared to the base case (no grid constraints), averaged over all scenarios.
    \item Standard deviation of the prosumers' cost increases from the base case as a percentage of the total cost increase. This is the metric of fairness.
    % \item Average non-zero voltage constraints violations in pu over all nodes and scenarios.
    \item Maximum voltage constraints violation in pu over all nodes and scenarios.
\end{enumerate}

% Grid constraints violations reflect the technical performance of the price signal methods. They are computed a posteriori, using the aggregated powers from the individual problem optimisations. In parallel, a cost analysis evaluates the impact of each price signal in comparison to the original cost of electricity, highlighting the trade-off between benefit and robust satisfaction of grid constraints.

\subsection{Comparison Between Expected and Robust Price Signals}

%\subsection{Tuning of $\rho$}
% The parameter $\rho$ represents the trade-off between cost-effectiveness and robustness to uncertainty. The evolution of the tradeoff is set as a Pareto front on a log-log graph. The parameter chosen as the best compromise corresponds to the elbow in the graph.

In this section, we compare the performance of the expected and robust price signals to assess whether they can respect the grid constraints while keeping the costs low. We use $\rho=10$ as the penalty on the grid constraints in the computation of the Lagrangian and ignore the fairness consideration, so $w^f=0$. The effect of $\rho$ and $w^f$ on the performance will be studied in the following section. The penalty $\alpha$ on the BESS power is set to $10^{-3}$.

\begin{figure}[t!]
    \centering
    \subfloat[]{
        \includegraphics[width=1\linewidth]{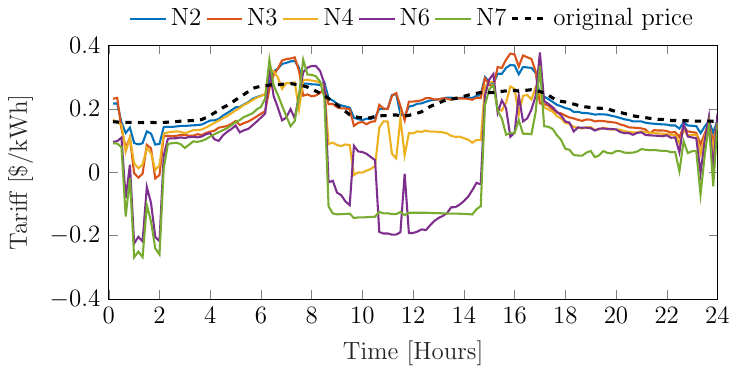}
    }
    \\
    \subfloat[]{
        \includegraphics[width=1\linewidth]{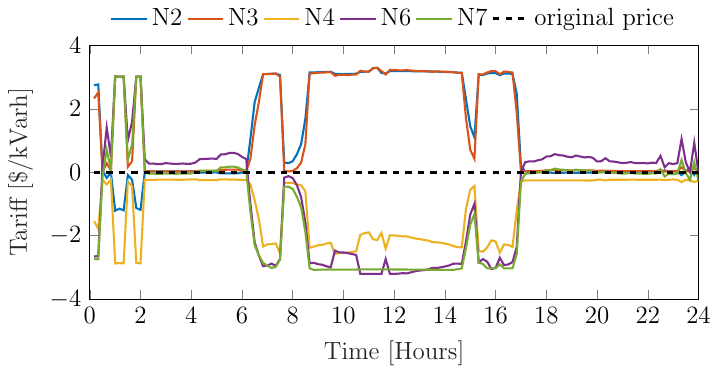}
    }
    \caption{Modified electricity tariffs for (a) active power and (b) reactive power for a single scenario.}
    \label{fig:price_signals}
\end{figure}

Fig.~\ref{fig:price_signals} shows the modified electricity tariffs for one scenario for which voltage constraints are binding (due to high PV generation and low loads). The original electricity price is the highest between 7~a.m. and 5~p.m., because that is the period when office buildings are the busiest. The price decreases during daytime, because of the high expected PV generation. The price computed by the DSO is lower than the original price at the beginning of the day, which incentivizes prosumers to charge their BESSs and sell their stored energy later when the price peaks. Selling the stored energy at this point offers a double benefit, as the prosumers benefit from the high price and can also absorb the incoming energy produced by the PVs. During the day, the prices advertised by the DSO are also low, so the prosumers can consume more. The price for prosumers at nodes N6 and N7, which are further away from the feeder, is negative, potentially resulting in PV curtailment. 

Examining the price for reactive power, we observe that the DSO can effectively compute incentives for prosumers to produce or consume reactive power in accordance with the grid's constraints. In particular, prosumers located farther from the feeder are incentivized to consume reactive power, thereby lowering the nodal voltages. In contrast, the other nodes need to produce reactive power to balance the production of the first prosumers.

\begin{figure}[t!]
    \centering
    \subfloat[]{
        \includegraphics[width=1\linewidth]{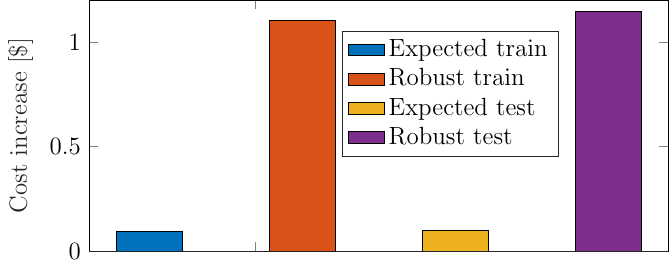}
    }\\
    \subfloat[]{ \label{fig:cost_std}
        \includegraphics[width=1\linewidth]{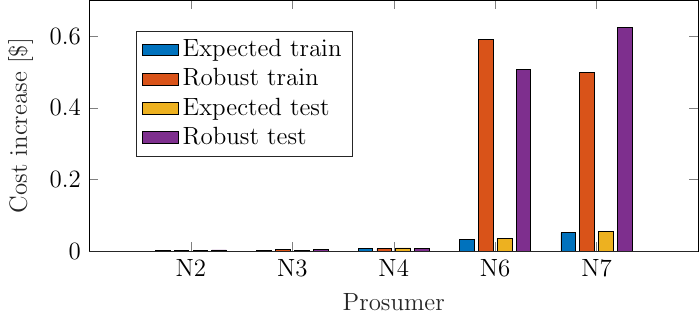}
    }\\
    \subfloat[]{
        \includegraphics[width=1\linewidth]{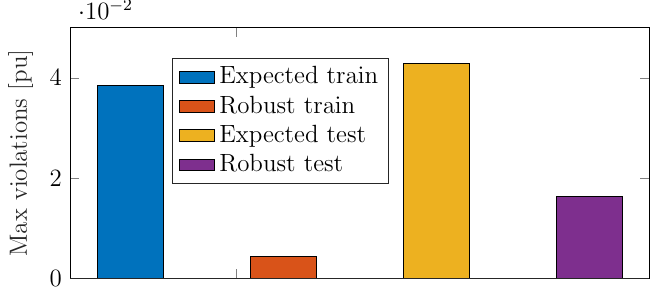}
    }
    \caption{Performance metrics for the expected and robust price signals evaluated on the training and test scenarios. (a) Total cost increase from the base case averaged over all scenarios, (b) average cost per prosumer, (c) maximum voltage violation over all scenarios.}
    \label{fig:metrics}
\end{figure}

% \begin{figure}[t!]
%     \centering
%     \subfloat[]{
%         \includegraphics[width=1\linewidth]{fig/mean_voltage.pdf}
%     }\\
%     \subfloat[]{
%         \includegraphics[width=1\linewidth]{fig/max_voltage.pdf}
%     }
%     \caption{(a) Average (non-zero) voltage violations and (b) maximum voltage violation for the expected and robust price signals evaluated on the training and test scenarios.}
%     \label{fig:voltage_violations}
% \end{figure}

Fig.~\ref{fig:metrics} shows the various performance metrics for the expected and robust price signals used on the training and test scenarios. The robust price signal results in a cost 10 times higher than the expected price signal. However, the robust price signal achieves significantly lower violations, which are below 1\% of the nominal value for the training scenarios. In contrast, the violations observed with the expected price signal constitute a significant percentage of the system's tolerance. Therefore, the cost increase imposed in the robust case is necessary to respect the voltage constraints. The maximum violation on the test scenarios is almost 3 times larger compared to the training scenarios. This implies that some of the worst-case scenarios have not been taken into account in the optimization problem. By considering more varied scenarios of the load profiles, we can prevent these worst cases.

Fig.~\ref{fig:cost_std} gives an insight into the fairness considerations of the computed price signals. We observe that, when fairness is not considered in the price signals, the prosumers at nodes N6 and N7 bear the overwhelming majority of the cost of the grid constraints. This is because these nodes have the most significant impact on the voltages, so it is more cost-effective to curtail the PVs of these prosumers to respect the voltage upper bound. Therefore, the DSO must consider fairness in the price signals it advertises to prosumers to avoid a single prosumer bearing the entire cost of the grid.

% Fig.~\ref{fig:voltage_violations} shows the average and maximum voltage violations on the grid over all nodes and scenarios while using the expected and robust price signals evaluated on the training and test scenarios.

\subsection{Effect of $\rho$ and Fairness}

\begin{figure}[t!]
    \centering
    \subfloat[]{\label{fig:rho_wf_train}
        \includegraphics[width=1\linewidth]{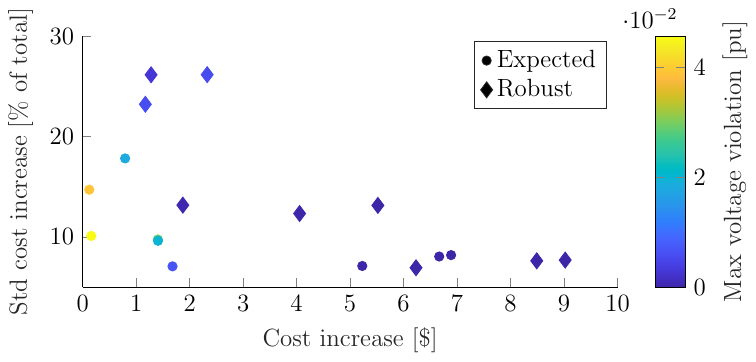}
    }\\
    \subfloat[]{
        \includegraphics[width=1\linewidth]{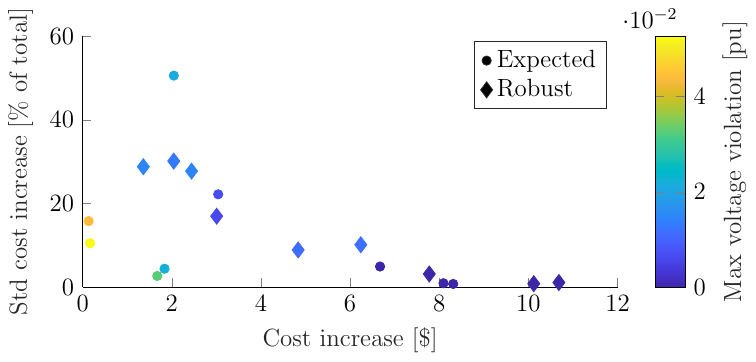}
    }
    \caption{Performance metrics for (a) training and (b) test scenarios for the robust and expected price signals. Each data point represents a different combination of the penalty $\rho$ on the grid constraints and fairness weight $w^f$.}
    \label{fig:rho_wf}
\end{figure}

In this section, we compute the performance metrics for different combinations of the penalty $\rho$ and the fairness weight $w^f$, in the range [1, 10, 1000]. Fig.~\ref{fig:rho_wf} represents the values of the cost increase, the standard deviation of cost increases (expressed as a percentage of the total cost), and the maximum voltage violations achieved by the expected and robust price signals for the training and test scenarios. First, we observe a trend of increasing fairness (decreasing variance of costs) as total costs increase. This is expected because equal cost-bearings can only be achieved by increasing the costs of the system. Moreover, the robust signal achieves worse fairness performance compared to the expected one for the same total cost; on the other hand, it achieves smaller voltage violations (the diamond points are darker shades of blue). This happens because the robust price signal considers the worst-case scenario, in which fairness might be sacrificed in favour of voltage control.

Looking at Fig.~\ref{fig:rho_wf_train}, we can make the following observations. If fairness is not necessary, which might be the case if the DSO remunerates the prosumers for their curtailment, then the robust price signal can achieve small average costs, around \$1-2 higher than the base case, while keeping the voltage violations smaller than 1\% of the nominal value. These costs can be achieved with a small fairness weight and a high value of $\rho$. On the other hand, although the expected price signal achieves lower costs, it fails to regulate the voltages in the worst-case scenario. If, however, fairness is desirable, the expected price signal can achieve similar performance in terms of voltage regulation and fairness, but with smaller average costs (\$5.2 instead of \$6.2) in the system, provided that both $\rho$ and $w^f$ are high. Therefore, the expected price signal can also achieve acceptable voltage regulation if paired with a high penalty $\rho$ on the grid constraints.

Similar observations can be made for the performance on the test scenarios. In this case, a good trade-off between fairness and costs can be achieved by both methods with an added average cost of \$8 and violations smaller than 1\%. However, the performance will depend on the similarity of the test scenarios to the training ones.

\section{Conclusion} \label{sec:conclusion}

We have proposed a method for computing price signals that enables prosumers to provide flexibility to the DSO, thereby respecting the distribution grid's constraints under uncertain operation. The method entails solving a scenario-based optimal power flow for different scenarios of the prosumers' demand and their PV generation, which also accounts for fairness in the costs experienced by the prosumers. 
%The DSO constructs a price signal for each prosumer and scenario using the augmented Lagrangian of the centralized problem, taking into account the grid's constraints and fairness in the costs among the prosumers. 
% We then compare two methods for aggregating the computed price signals, resulting in either an expected price signal for each prosumer, which only considers the expectancy of the loads and PV generation, or a robust one that selects the scenario with the highest potential constraint violations.
The DSO then computes a representative price signal for each prosumer that incorporates the grid's constraints and fairness.

We compare the performance of two methods for aggregating the computed price signals, resulting in an expected price signal and a robust one, using simulations in a model distribution grid.
% We compare the performance achieved by the two price signals in a model distribution grid using simulations. 
The validation assesses the ability of the two signals to adhere to the bounds of the nodal voltages while maintaining low costs and achieving a fair distribution of costs among prosumers. The expected price signal generally results in lower costs than the robust one, but fails to meet the required voltage constraints. Regarding fairness, both price signals can achieve a low variance of the prosumers' costs if the weights of the different objectives are carefully selected.

The scalability of the proposed method still requires thorough evaluation. Although the complexity of computing price signals scales linearly with the number of scenarios, the scenarios can scale exponentially with the number of nodes if all possible combinations among prosumers are considered. We should, therefore, carefully select the representative scenarios that account for the worst-case scenario of the system.

\bibliographystyle{IEEEtran}
\bibliography{references}

\end{document}